\documentclass[prd,a4paper,nofootinbib, 
showpacs,
]{revtex4}
\usepackage{graphicx}
\usepackage{axodraw}
\def\eq#1{{eq.~(\ref{#1})}}
\def\eqs#1#2{{eqs.~(\ref{#1})--(\ref{#2})}}
\def\fig#1{{fig.~(\ref{#1})}}
\def\vev#1{\left\langle #1\right\rangle}

\def\Tr{\mbox{Tr}\,}
\def\etal{{\it et al.}}

\def\ltap{\ \raisebox{-.4ex}{\rlap{$\sim$}} \raisebox{.4ex}{$<$}\ }

\def\hbar{\hspace{0pt}\raisebox{1pt}{$-$} \hspace{-7pt} h}

\def\5{\overline 5}

\newcommand{\be}{\begin{equation}}
\newcommand{\ee}{\end{equation}}
\newcommand{\bea}{\begin{eqnarray}}
\newcommand{\eea}{\end{eqnarray}}
\newcommand{\nn}{\nonumber}

\begin{document}
\title[The little flavons]{The little flavons
}
\date{June 14, 2003}
\author{F.~Bazzocchi}
\author{S.~Bertolini}
\author{M.~Fabbrichesi}
\affiliation{INFN, Sezione di Trieste and\\
Scuola Internazionale Superiore di Studi Avanzati\\
via Beirut 4, I-34014 Trieste, Italy}
\author{M.~Piai}
\affiliation{Department of Physics, Sloane Physics Laboratory\\
University of Yale, 217 Prospect Street\\
New Haven CT 06520-8120, USA}
\begin{abstract}

\noindent
Fermion masses and mixing matrices can be described in
terms of spontaneously broken (global or gauge) flavor symmetries.
We propose a little-Higgs inspired scenario in which an $SU(2)\times U(1)$
gauge flavor symmetry is spontaneously (and completely) broken by the vacuum of the dynamically
induced potential for  two scalar doublets (the flavons) which are pseudo-Goldstone bosons
remaining after the spontaneous breaking---at a scale between 10 and 100 TeV---of an approximate $SU(6)$ global symmetry.
 The vacuum expectation values of the flavons  give rise to the  texture in the fermion mass matrices.
We discuss in detail the case of leptons.  Light-neutrino masses arise
by means of a see-saw-like mechanism that takes place at the same
scale at which the $SU(6)$ global symmetry is broken.
We show that without any fine tuning of the parameters
the experimental values of the charged-lepton masses,
the neutrino square mass
differences and the Pontecorvo-Maki-Nakagawa-Sakata mixing matrix
are reproduced.
\end{abstract}
\maketitle
%
\vskip1.5em
\section{Motivations}

Goldstone bosons are  massless scalar particles remaining
after the
spontaneous breaking of global symmetries. Their number is determined by
the number of broken generators in the group algebra. Goldstone bosons
have no potential at all orders in perturbation theory
and only couple derivatively to other fields.
When  the global symmetry is explicitly broken, the
would-be-massless excitations acquire a potential and a mass
proportional to the
strength of the explicit breaking is generated.

When two or more independent global
symmetries are spontaneously broken by the same vacuum state,
only the simultaneous explicit breaking of all
symmetries lifts the flatness in the Goldstone boson potential.
This property has been recently exploited in the
\textit{little Higgs} models~\cite{littlehiggs} in order to stabilize against the one-loop quadratic renormalization
the scalar potential of the Higgs fields that is responsible for electroweak
symmetry breaking.
Explicit breaking is arranged in such a way that
more than one independent interaction term is needed in order
to break all of the spontaneously broken global symmetries (\textit{collective}
breaking).
As a result quadratic renormalization of the pseudo-Goldstone boson masses
arises only from the two-loop level.
This makes it possible
to increase by at least an order of magnitude the naturalness range of the
massive scalar theory.
In the standard model (SM) of electroweak interactions the natural
cut off is raised from few TeV's to few tens of TeV's.

This idea can be particularly attractive in a non-supersymmetric
context in which elementary
scalar fields are needed together with large scale differences between
their masses and the theory cut off.
%
%
A topical subject in which a scalar sector with hierarchical mass scales
appears and the naturalness issue arises is flavor physics.

The possibility that the  hierarchy between the masses of the SM fermions arises because of some global horizontal symmetry acting on the three generations of matter fields  has been extensively discussed in the literature~\cite{flavormodels-old, flavormodels-new}. In most of these models,  heavy scalar fields (referred to as \textit{flavons}) carry the quantum numbers of this flavor symmetry, and are responsible for its breaking by acquiring non-vanishing vacuum expectation values (VEV's). The Yukawa couplings
of the SM are then generated
from high-dimensional non-renormalizable operators which couple
fermions, Higgs fields and flavons.
The hierarchy between fermion masses is then the result of the
hierarchy between the VEV's and the cut-off scale of the theory which
controls the magnitude of all non-renormalizable operators.
Often, one is not interested in determining the scalar potential
of the flavon sector itself, and simply assumes the existence of these VEV's, with some hierarchical pattern deriving by
unknown UV properties and details of the
underlying theory.

What we propose here is to  obtain dynamically a stable (non-supersymmetric)
scalar potential by assuming that the flavons are pseudo-Goldstone
bosons originating from the breaking of an approximate global
symmetry, spontaneously broken to a subgroup containing the flavor
symmetry that acts on the SM fermions. In this way, the field content of the
flavon sector is determined.

The gauging of the flavor symmetry breaks explicitly the global symmetry
and induces a potential for the flavons.
The form of the potential as well the size of the scalar couplings
is obtained by means of the
Coleman-Weinberg potential~\cite{coleman-weinberg} of the
non-linear sigma model
describing the pseudo-Goldstone boson dynamics.

The potential induced by gauge interactions preserves the
$SU(2)_F\times U(1)_F$ flavor symmetry. The seed for the spontaneous breaking
of the flavor symmetry is given by (gauge invariant) interactions of
the two doublet flavons with  right-handed neutrinos.
These interaction terms destabilize the symmetric vacuum and drive the
complete breaking of the local flavor symmetry.
As a consequence all flavor mediating gauge bosons become massive.
At the same time, the one-loop stability of the flavon masses
on the broken vacuum is preserved.

While this approach is quite general, in this paper we focus on the lepton sector of the SM. The structure of the flavon interactions with the fermions
determines the lepton mass matrices and mixing. We show that all known
mass and mixing parameters are reproduced without any fine tuning
of the couplings of the model and that the needed patterns are derived
from the vacuum structure of the two doublet flavon potential.
We provide an explicit numerical example
that gives an instance of normal hierarchy for the neutrino
mass matrix and reproduces the Pontecorvo-Maki-Nakagawa-Sakata
(PMNS)~\cite{PMNS} mixing matrix as determined by the solar and
atmospheric experiments---as well as the experimental charged-lepton
masses. The same framework can be extended to the
quark sector and together with issues related to CP violation
will be the object of  further work~\cite{wip}.

The flavon masses are stabilized against gauge induced quadratic
renormalization at one-loop by a little-Higgs like mechanism~\cite{littlehiggs}
which will be described in detail in the next section.
The general framework is similar to that discussed in ref.~\cite{LSS}
and it actually uses the same high-energy global symmetry structure,
albeit with a different pattern of gauge symmetry breaking.
We have recalled this  similarity in the naming of the flavons.

\section{Goldstone bosons}

We choose as our basic flavor symmetry a gauged
$U(2)_F \simeq SU(2)_F\times U(1)_F$.
This choice is suggested by the approximate structure of the lepton sector:
both neutrinos and charged-leptons can be classified in first approximation
as two heavy flavor doublets---made by the $\tau$ and $\mu$ and the
corresponding neutrinos---and two lighter singlets---the $e$ and its neutrino.

To exploit the features of the little Higgs models, at least two copies of
the flavor group should be embedded in a larger (approximate) global symmetry.
The request that the flavon sector exhibits a vacuum structure that allows
for the complete breaking of the final gauge flavor symmetry is satisfied
minimally by two flavon doublets. The smallest group that satisfies these
requirements is $SU(6)$, spontaneously broken to $Sp(6)$, which has
been discussed as a little Higgs model in ref. \cite{LSS}.

In our model we assume that the electroweak and flavor
symmetries are embedded in two independent collective-breaking frameworks
with comparable cut-off scales $\Lambda_{H} \simeq \Lambda_{F} \simeq \Lambda$ between 10 and 100 TeV.
We will not enter the details of the ultraviolet completion of the model, and only
deal with the general structure of the effective theory below
the non-linear sigma model scale $f = \Lambda/4\pi$, where all the physics of flavor takes place.

Consider then the spontaneous breaking of a global flavor symmetry $SU(6)$
down to $Sp(6)$.  Fourteen of the generators of $SU(6)$ are broken
giving 14 (real) Goldstone bosons that can be written as a single field
\be
\Sigma = \exp \left( i \Pi/f \right)
\Sigma_0 \, .
\ee
They represent  fluctuations around the (anti-symmetric)
vacuum expectation value
\be
\Sigma_0 \equiv \langle \Sigma \rangle =
\left( \begin{array}{cc} 0 & -I \\ I & 0 \end{array} \right) \, .  \label{vacuum}
\ee

Within $SU(6)$ we can identify four subgroups as
\be
SU(6) \supset  [SU(2) \times  U(1)]^2 \, .
\ee
We choose to gauge these subgroups, in such a way as to explicitly break the
global symmetry through the gauge couplings.
Only the diagonal combination of these gauge groups
 survives the spontaneous breaking of the symmetry so that we have
\be
[SU(2) \times  U(1)]^2 \rightarrow SU(2) \times  U(1) \, .
\ee
We will use  the latter groups  to classify our fermion and flavon states.

The generators of the two $SU(2)$ are given by the $6\times6$ matrices
\be
Q_{1}^a = \frac{1}{2}\left( \begin{array}{cc|c} \sigma^a & 0 & 0\\
                                                 0 & 0 & 0\\
\hline
                                                    0 & 0 & 0

                                                     \end{array} \right)  \label{Q1}
\ee
and
\be
Q_{2}^a = \frac{1}{2}\left( \begin{array}{c|cc} 0 & 0 & 0\\
\hline
                                                 0 & - \sigma^{a*} & 0\\

                                                    0 & 0 & 0

                                                     \end{array} \right) \, , \label{Q2}
\ee
where $\sigma^a$ are the Pauli matrices;
we choose the  $U(1)$-charge matrices to be  given by
\bea
Y_1&= & -\frac{1}{2\sqrt{15}} \,\mbox{diag} \left[1,\,1,\,-5,\,1,\,1,1\, \right] \label{Y}\nn \\
Y_2&= & -\frac{1}{2\sqrt{15}} \,
\mbox{diag}\left[1,\,1,\,1,\,1,\,1,\,-5\, \right] \, . \eea

\begin{figure}
\begin{center}
\begin{picture}(260,100)(0,0)
\LongArrow(110,20)(150,20)
\LongArrow(80,80)(170,80)
\LongArrow(40,70)(40,30)
\LongArrow(220,70)(220,30)
\Text(40,20)[c]{$\left[SU(2)\,\times\, U(1)\right]^2\,\times\,U(1)_P$}
\Text(220,20)[c]{$\left[SU(2)\,\times\, U(1)\right]\,\times\,U(1)_P$}
\Text(40,80)[c]{$SU(6)$}
\Text(220,80)[c]{$Sp(6)$}
\end{picture}
\end{center}
\caption{Diagrammatic representation of the symmetry structure of the sigma model.
Horizontal arrows indicate the spontaneous $SU(6)\to Sp(6)$
global symmetry breaking, vertical arrows
the explicit breaking due to gauge interactions.
A global $U(1)_P$ is preserved both by
the spontaneous and the explicit breaking (induced by gauge interactions)
while it is
explicitly broken by the Yukawa sector of the model (see text). }
\label{fig:symmetries}
\end{figure}
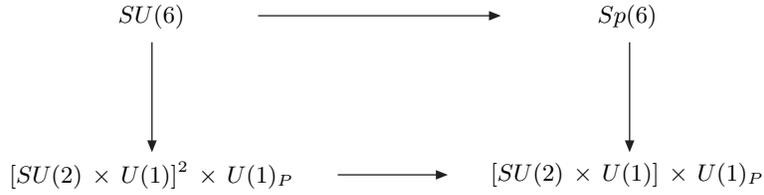

Contrary to \cite{LSS}, our  $U(1)$ charges belong to the generators of the group $SU(6)$.
Notice that the gauged subgroup $[SU(2)\times U(1)]^2$ has rank 4:  this means that one of the generators
of the Cartan sub-algebra of $SU(6)$ (rank 5) is neither gauged nor explicitly broken.
We identify this generator with
\bea
P\,=\,\mbox{diag} \left[1,\,1,\,0,\,-1,\,-1,0\, \right] \,,
\eea
in such a way that it commutes with the whole gauge group $\left[SU(2)\,\times\,U(1)\right]^2$,
and it is orthogonal to all its generators. The generator $P$ belongs
to the algebra of $Sp(6)$, and  generates a $U(1)_P$ exact global symmetry of the sigma model we are discussing.
This symmetry is then explicitly broken by the
couplings of flavons to fermions, as we shall discuss.
We summarize the symmetry structure of the sigma model in Fig~\ref{fig:symmetries}.

In the low-energy limit, there are two scalar bosons,
\be
\phi_1 = { \phi_1^+  \choose \phi_1^0} \quad \mbox{and}\quad   \phi_2 = {\phi_2^0 \choose \phi_2^-} \, ,
\ee
 that are $SU(2)$-doublets  with $U(1)$ charges respectively $1/2$ and $-1/2$,
 and one $SU(2)$- and $U(1)$- singlet $s$. The remaining four bosons are eaten in the breaking of the (gauge) $[SU(2) \times  U(1)]^2$ symmetries.
Accordingly we find that in the low-energy limit we can write the pseudo-Goldstone boson matrix as
\be
\Pi = \left( \begin{array}{cccccc} 0& 0& \phi_1^+ & 0 & s & \phi_2^0 \\
                                                    0& 0& \phi_1^0 &  -s & 0 & \phi_2^- \\
                                                     \phi_1^- & \phi_1^{0*} &0 & -\phi_2^0 & -\phi_2^- &0 \\
                                                     0 & -s^* & -\phi_2^{0*} & 0 & 0 & \phi_1^{-}\\
                                                     s^* & 0 & -\phi_2^+ & 0 & 0 & \phi_1^{0*}\\
                                                     \phi_2^{0*} & \phi_2^+ & 0 & \phi_1^+ & \phi_1^0 & 0
                                                     \end{array} \right) \, .
\ee
The singlet field becomes massive and has no expectation value in the vacuum configuration we will use; it is therefore effectively decoupled from the theory. The two doublets are our \textit{little flavons}.

Under the action of $U(1)_P$ global transformations $U\,=\exp \, (i\,\alpha P)$,
the doublets and singlet transform as:
\be
\phi_{1,2}  \longrightarrow  e^{i\,\alpha}\,\phi_{1,2}\ , \quad
s  \longrightarrow  e^{2\,i\,\alpha}\,s\,.
\ee

By construction, all Goldstone bosons start out massless and with only
derivative couplings. However, as anticipated, the gauge
and flavon-fermion interactions explicitly
break the symmetry and give rise to an effective potential for the
pseudo-Goldstones, the form of which allows for the
existence of a non-symmetric vacuum that
completely breaks the residual flavor gauge symmetry.

\section{The effective potential}

The effective lagrangian of the pseudo-Goldstone bosons below the symmetry-breaking scale  is given by the kinetic term
\be - \frac{f^2}{4} \Tr
\left( D^\mu \Sigma \right) \left( D_\mu \Sigma \right)^\ast\, ,
\label{kinetic}
\ee
where the minus sign follows from the antisymmetric form of $\Sigma$.
As already mentioned, we take the cut-off scale
$
\Lambda = 4\, \pi f
$
to be of the order of 10-100 TeV.
The absolute value of this scale is immaterial to the generation
of the lepton mass matrices that, as we shall see,
only depend on the ratio between the
vacuum expectation values, which are proportional to $f$, and $f$
itself. On the other hand, too large a scale would
destabilize the standard model Higgs mass via interactions with the flavons
and introduce a fine-tuning (we comment on this issue in Sect. IV).

The covariant derivative in (\ref{kinetic}) is given by
\be
D_\mu \Sigma = \partial_\mu + i g_i A^a_{i_{\mu}} \left( Q_i^a \Sigma + \Sigma Q_i^{aT} \right)+ i g'_i B_{i_\mu} \left( Y_i \Sigma + \Sigma Y_i^T \right)
\ee
where $A^a_{i_{\mu}}$\, and\, $B_{i_\mu}$\, are the gauge bosons of the $SU(2)_i$\, and \, $U(1)_i$ gauge groups respectively and $Q_i^a$\, and \,$Y_i$ their generators as given in (\ref{Q1}), (\ref{Q2}) and (\ref{Y}). Since the vacuum $\Sigma_0$ in \eq{vacuum} breaks the symmetry $(SU(2) \times U(1))^2$ into the diagonal $SU(2) \times U(1)$, four combinations of the initial gauge bosons become massive; their masses are given by
\be
M^2_{A'} = \frac{(g_1^2 +g_2^2)f^2}{2} \quad \mbox{and} \quad M^2_{B'} = \frac{(g_1^{'2} +g_2^{'2})f^2}{4} \,. \label{MV}
\ee

The effective potential must break the $SU(2)\times U(1)$ remaining gauge symmetry and give mass to all surviving pseudo-Goldstone bosons, little flavons included.
At one loop, the gauge interactions give rise to the Coleman-Weinberg potential given by the two terms
\be
\frac{\Lambda^2}{16\pi ^2}\Tr [M^2(\Sigma)] + \frac{3}{64 \pi^2}\Tr \left[M^4(\Sigma)\left(\log\frac{M^2(\Sigma)}{\Lambda^2} + {\rm const.} \right)\right] \, .
\label{CW}
\ee
In agreement with the general framework of  little Higgs models  the quadratically divergent term gives mass only to the singlet fields $s$. No mass is generated for the (doublet) little flavons.  In addition, a trilinear coupling between the doublets $\phi_1$ and $\phi_2$ and $s$  and  a quartic term for the two doublets are generated:
\be
\frac{\Lambda^2}{16\pi ^2}\Tr [M^2(\Sigma)]= f^2\left(3\,g_1^2\left| s + \frac{i}{2f}\tilde{\phi_2}^\dag \phi_1\right|^2 + 3\,g_2^2\left| s - \frac{i}{2f}\tilde{\phi_2}^\dag \phi_1\right|^2  \right) \, ,
\label{quad}
\ee
where $\tilde{\phi} = i \sigma_2 \phi^*$.
From \eq{quad} one obtains
\be
m_s^2 = \frac{3}{2} (g_1^2 +g_2^2)f^2 \, ,
\ee
and the quartic coupling
\be
\lambda_4 \,  |\tilde{\phi_2}^\dag \phi_1|^2 \, .
\label{L4term}
\ee
After integrating out  the heavy singlet $s$, one obtains for $\lambda_4$ the
cut-off independent expression:
\be
\lambda_4 = \frac{g_1^2g_2^2}{g_1^2 +g_2^2} \simeq O(g^2)\, ,
\label{L4}
\ee
which is the only term generated by the quadratic term in (\ref{CW}). This happens
because of  the mechanism of collective breaking  for which the potential of the
pseudo-Goldstone boson doublets (little flavons)
is generated  by the interplay of both gauge interactions, thus breaking explicitly the global $SU(6)$ symmetry,
while at the same time protecting the doublets from receiving a (quadratically
divergent) mass at the one-loop level.

Mass terms as well as other effective quartic couplings for the little flavons arise from the logarithmically divergent term in \eq{CW}.
One can verify that
the one-loop potential induced by gauge interactions includes
the following terms
\be
 \mu_1^2 \phi_1^\dag \phi_1 + \mu^2_2 \phi_2^\dag \phi_2+ \lambda_1 (\phi_1^\dag \phi_1)^2 + \lambda_2 (\phi_2^\dag \phi_2)^2
 +   \lambda_3 (\phi_1^\dag \phi_1)(\phi_2^\dag \phi_2) \, .
\label{logterms}
\ee
The size of the  mass terms and effective couplings are given by
\be
\mu_{i}^2/f^2 \simeq \lambda_i  \simeq c{(\mu_i,\lambda_i)} \frac{3g^4}{64 \pi^2}
\log \frac{M^2_V}{\Lambda^2} \ltap 10^{-2} \, ,
\label{M4}
\ee
where $c$ are numerical coefficients, related to the  expansion of the $\Sigma$, and $M_V$ is the mass of the massive gauge bosons (see, \eq{MV}).
The numerical estimate in \eq{M4} takes into account that we take the (horizontal) gauge symmetry coupling to be of $O(1)$. As we shall see,
this follows from requiring the
cut off of the model to be as low as possible (at a scale comparable
with the SM little Higgs cut off) while avoiding too light flavons.
In particular, we entertain the possibility of a flavor interaction cut off
$\Lambda \simeq 10-100$ TeV and flavon masses just above the weak scale.

Since $\lambda_4$, induced by the leading divergent term
in the Coleman-Weinberg potential, turns out to be a sizeable coupling,
other relevant contributions to the effective potential may arise from
integration of the doublet self-interaction in \eq{L4term} which contributes
to the $\lambda_{1,2}$ terms
with
\be
\lambda_{1, 2} \simeq \frac{\lambda_4^2}{64 \pi^2}
\log \frac{\Lambda^2}{M_\phi^2}  \ltap 10^{-2}\, ,
 \label{L1L2loopL4}
\ee
which are in fact of the same order of those induced by the logarithmically
divergent term in the gauge induced one-loop effective potential.

The one-loop flavon potential generated by gauge interactions in
\eqs{L4term}{logterms} has to be compared with the
general potential for two $SU(2)$ doublets of opposite hypercharges that is
given, up to four powers of the fields,  by
\bea
V_4(\phi_1, \phi_2)  & = &  \mu_1^2 \phi_1^\dag \phi_1 + \mu^2_2 \phi_2^\dag \phi_2
+ (\mu_3^2 \tilde \phi_1^\dag \phi_2 + H.c.)
\nn \\
 & +&   \lambda_1 (\phi_1^\dag \phi_1)^2 + \lambda_2 (\phi_2^\dag \phi_2)^2
 +   \lambda_3 (\phi_1^\dag \phi_1)(\phi_2^\dag \phi_2) + \lambda_4 |\tilde \phi_1^\dag \phi_2|^2
  + \lambda_5 \big[ (\tilde{\phi_1}^\dag \phi_2)^2 + H.c.
\big]
\label{genpot}
\eea

Depending on the sign of the determinant of the mass matrix of the scalar
fields and on relationships among the various couplings,
the potential in \eq{genpot} can have different symmetry breaking
minima~\cite{sher}. In particular we are interested to the vacuum which completely breaks
the $SU(2)\times U(1)$ gauge flavor symmetry.
The residual exact $U(1)_P$ global symmetry, acting with opposite charge on $\phi_i$ and $\tilde{\phi}_i$ fields, forbids the generation of the $\mu_3$ and $\lambda_5$ couplings.
In the absence of $\mu_3^2$ and $\lambda_5$
terms,  the vacuum can be parametrized as
\be
\langle \phi_1 \rangle = { 0  \choose v_1} \quad \quad \langle \phi_2 \rangle = {0 \choose v_2}
\label{vacuum2}
\ee
with real VEV's.
The complete breaking of the flavor symmetry
allows us to avoid the presence in the physical spectrum of
massless flavor gauge bosons and is necessary in order
to generate the lepton mass matrices.
This vacuum breaks also the $U(1)_P$ symmetry,
however a linear combination of
$P$ and flavor isospin is still preserved.
The corresponding global symmetry $U(1)_{P'}$, with
\bea
P^{\prime}\,=\,{\rm diag}\,\left[1,\,0,\,0,\,-1,\,0,\,0\,\right]\, ,
\eea
is explicitly broken by the Yukawa sector.

The requirement that the potential is bounded from below gives the three conditions
\be
\lambda_1 + \lambda_2 > 0,
\quad
4 \lambda_1\lambda_2 -\lambda_3^2>0,
\quad
\mbox{and}
\quad
\lambda_4 - |\lambda_5| > 0
\label{bounded} \, .
\ee
Assuming $\mu_{1,2}^2 < 0$ (and making use of  $\mu_3^2 = \lambda_5=0$)
the symmetry breaking vacuum in \eq{vacuum2} leads to
the following flavon mass spectrum
\bea
m_{1,2}^2 & = & m_{5,6}^2  =  0\nonumber \\
m_{3,4}^2 & = & \frac{1}{2}  \lambda_4 \left( v_1^2 + v_2^2  \right) \nonumber \\
 m_{7,8}^2 &=& \lambda_1 v_1^2 + \lambda_2 v_2^2 \pm \sqrt{ (\lambda_1 v_1^2 + \lambda_2 v_2^2)^2 -( 4 \lambda_1\lambda_2 -\lambda_3^2) v_1^2 v_2^2}
 .
\label{gmass}
\eea
Positivity of the mass eigenvalues then requires
$ \lambda_4 > 0 $ and $\lambda_1 v_1^2 + \lambda_2 v_2^2> 0$.

The four massless degrees of freedom  are eaten by the four gauge fields
of the completely broken $SU(2)\times U(1)$ flavor symmetry
which become massive at a scale determined by the VEV's size
\be
v_1^2 = - \frac{2 \lambda_2\mu_1^2 - \lambda_3\mu_2^2}{4\, \lambda_1\lambda_2
-\lambda_3^2} \quad
v_2^2 = - \frac{2 \lambda_1\mu_2^2 - \lambda_3\mu_1^2}{4\, \lambda_1\lambda_2 -\lambda_3^2} \, .
\label{VEVs}
\ee
The condition $\mu_1^2,\ \mu_2^2 <0$ can be realized if there exist
fermions coupled to the doublets that
induce contributions to the scalar masses of opposite sign with respect to that
induced by gauge interactions.
This role is played in the model by heavy right-handed Majorana neutrinos,
with mass $M\simeq f$.
Radiative contributions to the flavon potential arising from global
$SU(6)$ breaking couplings to Majorana right-handed neutrinos
(as given in the next section) lead to scalar mass terms
\be
\mu_{1, 2}^2 \simeq - c^{(1,2)}_{n} \, \eta_{n}\ \frac{\Lambda^2}{16\pi^2}
\simeq - c^{(1,2)}_{n}\,  \eta_{n}\ f^2 \, ,
\label{mu12eta}
\ee
where $c^{(1,2)}_{n}$ are coefficients of order unity.
These quadratically-divergent corrections maintain the flavon mass
scale below the $f$ scale (and in the TeV regime) as long as $\eta_i \ltap 10^{-2}$.
Thus, still avoiding a large fine tuning of the couplings, no collective breaking
mechanism is required for the lepton-induced renormalization
(the only large couplings in the model are gauge and the Yukawa of the
top quark).

Notice that, the contributions to the quartic couplings induced by the massive right-handed
neutrinos are therefore given by
\be
\lambda_{1,2,3} \simeq \frac{c^{(1,2,3)}_{nm} \eta_{n}\eta_{m}}{16\pi^2}
\log \frac{\Lambda^2}{M^2} \ltap 10^{-6}
\label{loopn}
\ee
and are subleading, with respect to those induced by gauge interactions.

From \eq{VEVs} and \eqs{M4}{mu12eta} we obtain $v_1,\ v_2 = O(f)$,
which in turn implies that the four flavor gauge bosons and two of the flavon states have masses of order $f$, while the remaining two
scalars have masses of $O(10^{-1} f)$.
As anticipated, considering the lightest flavon states to be in the weak scale range
puts the flavor cut off $\Lambda = 4\pi f$ in the $10-100$ TeV regime,
in the same ballpark of the SM little-Higgs cut off.

Assuming all of the above conditions satisfied (we will not be concerned with
the detail of the ultraviolet completion of the theory) we now discuss
the neutrino and charged lepton mass textures that arise by assigning
non-trivial flavor transformation properties to the lepton families.

 \section{Coupling  flavons to  leptons}

We classify leptons of different families according to the
$[SU(2)\times U(1)]^2$ gauge flavor symmetry. As we have seen, the spontaneous breaking of
the global $SU(6)\rightarrow Sp(6)$ (approximate) symmetries
leads to the $[SU(2) \times U(1)]^2 \rightarrow SU(2) \times U(1)$ breaking.
We let all fermions to transform
under only one of the initial $SU(2)\times U(1)$ groups, so that
their charges will coincide with those of the surviving
diagonal group. This choice  determines the possible terms appearing
in the flavon interactions.
We will indicate the remaining $SU(2)\times U(1)$ flavor symmetry
with the index
$F$ to distinguish it from the electroweak group. In the following, all Greek
indices belong to the flavor group while Latin indices refer to the
electroweak group.
We assume all the fermions to be neutral under $U(1)_P$.

The standard model electron doublet ${l}_{e L}$ is an $SU(2)_{F}$
singlet charged under $U(1)_{F}$, while ${l}_{\mu,\tau L}$
are members of a doublet, that is
\be
{l}_{1 L}={l}_{e L}\ , \quad Y_{F}=- 2\,;\,\quad L_{L} =
\left(l_{\mu}\,,\,l_{\tau} \right)_L, \quad Y_{F} = \frac{1}{2}\, ;
\ee
where we have written explicitly the flavor hypercharge.
Right-handed charged leptons have a similar structure
\be
e_{R}\ , \quad Y_{F}= 1;\,\quad E_{R} = \left( \mu\, ,\,\tau \right)_R\ ,
\quad Y_{F}= \frac{1}{2} \, .
\ee

In order to have a \textit{see-saw}-like  mechanism~\cite{seesaw}, we introduce three right-handed neutrino $\nu^{i}_{R}$  which are $SU(2)_{F}$ singlets :
\be
\nu_{1 R}\ , \quad Y_{F}= 1\ ; \quad \nu_{2 R}\ , \quad Y_{F}= - 1\ ; \quad \nu_{3 R}\ , \quad Y_{F}= 0 \, .
\ee
This choice allows us to have  in the effective lagrangian Majorana mass entries at the scale  $M\sim f$.

\begin{table}[ht]
\begin{center}
\caption{Summary of the charges of all leptons and the relevant component of the pseudo-Goldstone bosons under the horizontal flavor groups $SU(2)_F$ and $U(1)_F$.}
\label{fields}
\vspace{0.2cm}
\begin{tabular}{|c|c|c|}
\hline
 $\alpha = 2, 3 $ &\quad\quad $U(1)_F$ \quad\quad &\quad $SU(2)_F$ \quad \cr
\hline
${l}_{e L}$ &   $-2$ & 1 \cr
$e_R$  & $1$ & 1 \cr
${L}_L = ( {l_{\mu}}\, ,\,{l_{\tau}})_L$  & $1/2$ & 2  \cr
$E_R = (\mu \,,\,\tau)_R$  & $1/2$ & 2  \cr
$\nu_{1R}$  & $1$ & 1  \cr
$\nu_{2R}$  & $-1$ & 1  \cr
$\nu_{3R}$  & $0$ & 1  \\[1ex]
\hline
$\Sigma_{\alpha-1\, 6} = (- i/f\ \phi_1 + ...)_{\alpha}$ &1/2 & 2 \\
$\Sigma_{\alpha-1\, 3} = (+ i/f\ \phi_2 + ...)_{\alpha}$ &$-1/2$ & 2 \\
$\Sigma_{3\, 2+\alpha} = (- i/f\ \phi_1^* + ...)_{\alpha}$ & $-1/2$&  $2^*$ \\
$\Sigma_{6\, 2+\alpha} = (- i/f\ \phi_2^* + ...)_{\alpha}$ &1/2 &  $2^*$ \\
\hline
\end{tabular}
\end{center}
\end{table}

The physical PMNS mixing is generated thanks to the different assignments
of right-handed charged leptons and neutrinos, which give different
textures to the charged lepton mass matrix,  the Dirac mass matrix of neutrinos
and  the (non-trivial) Majorana mass matrix of right-handed  neutrinos.

Taking into account the charge assignments, as summarized in Table I,
we construct the Yukawa interactions by  coupling, in a gauge invariant manner, the leptons to the pseudo-Goldstone bosons $\Sigma$ and to the standard-model Higgs boson. At the first non-trivial order in powers of the $\Sigma$ fields
we have:
\bea
\mathcal{L}_{\nu}  &=&
\left[ \lambda_{1\nu}\,\overline{\nu_{1R}}\ \left( \Sigma_{\alpha-1\,6}\Sigma_{6\,2+\alpha} \right)^{-Y_{1L}+ Y_{\nu_{1R}}} \right.
\ +\  \lambda_{2\nu}\,\overline{\nu_{2R}}\ \left( \Sigma_{\alpha-1\,6}\Sigma_{6\,2+\alpha} \right)^{-Y_{1L}+ Y_{\nu_{2R}}} \nn \\
&&\ + \left. \lambda_{3\nu}\,\overline{\nu_{3R}}\ \left( \Sigma_{\alpha-1\,6}\Sigma_{6\,2+\alpha} \right)^{-Y_{1L}+ Y_{\nu_{3R}}}\right]
(\tilde{H}^\dag\ {l}_{1 L} ) \nn \\
&+& i\ \Big[ \overline{\nu_{1R}}\ \left( \lambda'_{1\nu}\,\epsilon_{\alpha\beta} \Sigma_{\beta-1\,6}
+ \lambda''_{1\nu}\,\Sigma_{6\,2+\alpha} \right)
\ +\ \overline{\nu_{2R}}\ \left( \lambda'_{2\nu}\,\epsilon_{\alpha\beta} \Sigma_{\beta-1\,3}
+  \lambda''_{2\nu}\,\Sigma_{3\,2+\alpha} \right)\
(\Sigma_{\delta-1\,6}\Sigma_{6\,2+\delta}) \nn \\
&&\quad +\ \overline{\nu_{3R}}\ \left( \lambda'_{3\nu}\,\epsilon_{\alpha\beta} \Sigma_{\beta-1\,3}
+ \lambda''_{3\nu}\,\Sigma_{3\,2+\alpha} \right) \Big]
(\tilde{H}^\dag\ {l}_{\alpha L} ) \nn \\
&+& \left(-\frac{M}{2} + \frac{\eta_1 f}{2}\; \Sigma_{\alpha-1\,6}\Sigma_{3 \, 2+\alpha} + \frac{\eta_2 f}{2}\; \Sigma_{\alpha-1\,3}\Sigma_{6\,2+\alpha}\right)
\left(\overline{\nu^c_{1R}} \nu_{2R} + \overline{\nu^c_{2R}} \nu_{1R}\right)
\nn \\
&+& \left(-\frac{M_3}{2} + \frac{\eta_3 f}{2}\; \Sigma_{\alpha-1\,6}\Sigma_{3 \, 2+\alpha} + \frac{\eta_4 f}{2}\; \Sigma_{\alpha-1\,3}\Sigma_{6\,2+\alpha}\right) \left(\overline{\nu^c_{3R}} \nu_{3R} \right) \nn \\
&+&  \frac{\eta_{5}f}{2}\ ( \Sigma_{\alpha-1\,6}\Sigma_{6\,2+\alpha})^{\dag 2}\; \overline{\nu^c_{1R}} \nu_{1R} \,+\, \frac{\eta_{6}f}{2}\ ( \Sigma_{\alpha-1\,6}\Sigma_{6\,2+\alpha})^2\;
\overline{\nu^c_{2R}} \nu_{2R}  \\
&+& \frac{\eta_{7}f}{2}\  ( \Sigma_{\alpha-1\,6}\Sigma_{6\,2+\alpha})^\dag
\left(\overline{\nu^c_{1R}} \nu_{3R} + \overline{\nu^c_{3R}} \nu_{1R} \right)
 +\frac{\eta_{8}f}{2} \;  ( \Sigma_{\alpha-1\,6}\Sigma_{6\,2+\alpha})  \left(\overline{\nu^c_{2R}} \nu_{3R} + \overline{\nu^c_{3R}} \nu_{2R}\right) + H.c. \, ,\nn
\label{yuknu}
\eea
where $\epsilon = i \sigma_2$ is the completely antisymmetric tensor  with
indices $ \alpha,\beta = 2,\, 3 $ ($2 \rightarrow \mu ,\ 3 \rightarrow \tau$),
$\tilde{H} = i \sigma_{2} H^*$, and $\psi^c = C \bar \psi^T$.

Analogously, for the charged leptons we have 
\bea \mathcal{L}_{e}
&=& \overline{e_{R}}\ \left[ \lambda_{1e}\
(\Sigma_{\alpha-1\,6}\Sigma_{6\,2+\alpha})^{(-Y_{1L}+Y_{1R})}\
(H^\dag\ {l}_{1L})
+ i\ (\lambda_{3e}\  \Sigma_{6\,2+\alpha} + \lambda_{2e}\  \epsilon_{\alpha\,\beta} \Sigma_{\beta-1\,6})
(H^\dag\ {l}_{\alpha L}) \right]  \nn \\
&+&  \overline{E_{\alpha R}}\Big[ i
\left( \lambda_{1E}'\  \Sigma_{6\,2+\alpha} + \lambda_{1E}\ \epsilon_{\alpha\,\beta} \Sigma_{\beta -1\,6} \right)
(\Sigma_{\delta-1\,6}\Sigma_{6\,2+\delta})^{-Y_{1L}} \Big]
(H^\dag\ {l}_{1L})  \nn \\
&+&  \overline{E_{\alpha R}}\Big[
\delta_{\alpha\,\beta} \big(
- \lambda_{2E}\ + \lambda_{2E}'\ \Sigma_{\gamma-1\,6}\Sigma_{3\,2+\gamma}
+ \lambda_{2E}''\  \Sigma_{\gamma-1\,3}\Sigma_{6\,2+\gamma} \big)
  \nn \\
&& \quad\quad\  
+ \Big(\lambda_{3E}\  \Sigma_{\alpha-1\,6}\Sigma_{3\,2+\beta}
+ \lambda_{3E}' \ \epsilon_{\alpha\,\delta}\epsilon_{\beta\,\gamma}\ \Sigma_{6\,2+\delta}\Sigma_{\gamma-1\,3} + (3\leftrightarrow 6) \Big) \nn \\
&& \quad\quad\
+\Big( \lambda_{4E}\  \Sigma_{\alpha-1\,6}\Sigma_{\gamma-1\,3}
\epsilon_{\beta\,\gamma}
+ \lambda_{4E}'\ \epsilon_{\alpha\,\delta}
\Sigma_{6, 2+\delta}\Sigma_{3, 2+\alpha}+ (3\leftrightarrow 6)
\Big) 
\Big] (H^\dag\ {l}_{\beta L})\  + H.c. \, .
\label{yuklep}
\eea

In \eqs{yuknu}{yuklep} we have not included terms $\epsilon_{\gamma\,\delta} \Sigma_{\gamma-1\,6}\Sigma_{\delta-1\,3} = \tilde \phi_1^\dag \phi_2/f^2 + ...$
which vanish on the vacuum of \eq{vacuum2} and do not contribute to the
fermion masses.

The $U(1)_P$ symmetry
is explicitly broken by the lepton Yukawa couplings. The leading breaking
is due to the terms linear in $\tilde\phi_1$ and $\phi_2$ which arise from
the terms linear in the $\Sigma$ fields
in \eq{yuklep}.
On the other hand, these $U(1)_P$ violating terms have a negligible
impact on the
flavon potential: as it will be clear from the section devoted to the
numerical discussion of the mass matrices and mixing,
none of the Yukawa parameters entering the lagrangian is large (typically they are
of the order of $10^{-2}$ or less).  As a consequence, the $\mu_3^2$ terms---quadratically divergent contributions notwithstanding (two-loops are needed
above the electroweak symmetry breaking scale)---turn out to be negligible with
respect to $\mu_{1,2}^2$.

Analogously flavon-Higgs mixing terms of the type
$H^\dag H \phi_i^\dag \phi_i$---induced at one-loop by combinations of lepton
Yukawa couplings---do not destabilize the electroweak breaking as long as
we consider flavon vacuum expectation values below 10 TeV together
with Yukawa couplings of the order of $10^{-2}$ or less.

In our approach right-handed neutrinos are integrated out at the
$f$ scale thus realizing
a low-energy see-saw mechanism (similar realizations have been recently proposed in the context of extended technicolor~\cite{tom} and
deconstruction~\cite{lindner}).
The Dirac and Majorana matrices combine yielding below the $f$ scale
effective Yukawa couplings of the form:
\be
(\overline{{l}^c_{\rho L} }\ \tilde H^*)(\tilde H^\dag\ {l}_{\tau L})
\left[{\cal M}^T_{RL}(\Sigma) {\cal M}^{-1}_{RR}(\Sigma) {\cal M}_{RL}(\Sigma)
\right]_{\rho\,\tau} \label{ss}
\ee
where ${l}^c_{\tau L} = (\nu^{c}_\tau , e^{c}_\tau )_{L}$ and
$\tau,\ \rho = 1,2,3$.
Taking the leading non-vanishing orders in ${\cal M}^{-1}_{RR}(\Sigma)$ and
in the number of $\Sigma$ fields we obtain:
\bea
-2\, \mathcal{L}_{\nu} & = &
\frac{(\overline{{l}^c_{1L} }\ \tilde H^*)(\tilde H^\dag\ {l}_{1L})}{M}\
\left[2 \lambda_{1\nu} \lambda_{2\nu} + r\ \lambda_{3\nu}^2 \right] \left[\Sigma_{\alpha-1\,6}\Sigma_{6\,2+\alpha} \right]^{-2 Y_{1L}} \nn \\
& + &
 \frac{ (\overline{{l}^c_{1L} }\ \tilde H^*)(\tilde H^\dag\ {l}_{\alpha L}) + (\overline{{l}^c_{\alpha L} }\ \tilde H^*)(\tilde H^\dag\ {l}_{1L}) }{M}
 \lambda_{2\nu} (\lambda_{1\nu}' \ \epsilon_{\alpha\beta} \Sigma_{\beta-1\,6}
+ \lambda_{1\nu}''\ \Sigma_{6\,2+\alpha})
  \left[\Sigma_{\delta-1\,6}\Sigma_{6\,2+\delta} \right]^{-Y_{1L}+Y_{\nu_{2R}}}   \nn \\
&+ &
\frac{(\overline{{l}^c_{\alpha L} }\ \tilde H^*)(\tilde H^\dag\ {l}_{\beta L})}{2 M_3}
(i\sigma_2 \sigma_{\tau})_{\alpha\, \beta}\  (i\sigma_2 \sigma_{\tau})_{\delta\,\gamma}
\left[(\lambda_{3\nu}')^2 \ \Sigma_{\delta-1\,3}\Sigma_{\gamma-1 \,3}
\right.  \nn \\
&+ &  \left.
 \lambda_{3\nu}' \lambda_{3\nu}'' \ ( \epsilon_{\gamma\gamma'} \Sigma_{\delta-1\,3}\Sigma_{3\,2+\gamma'} + \delta\leftrightarrow\gamma )
+(\lambda_{3\nu}'')^2\ \epsilon_{\delta\delta'}\epsilon_{\gamma\gamma'}\Sigma_{3\,2+\delta'} \Sigma_{3\,2+\gamma'} \right] + H.c. \, , 
\eea
where $r = M/M_3$,  $\sigma_{\tau}/2$ are the generators of  the
$SU(2)_f$ gauge group ($\tau=1,2,3$).

At the leading order in the pseudo-Goldstone boson expansion the effective
Yukawa lagrangian is given by
\bea
-2\, \mathcal{L}_{\nu}
& =&  \left[2 \lambda_{1\nu} \lambda_{2\nu} + r\ \lambda_{3\nu}^2 \right]
\frac{(\overline{{l}^c_{1L} }\ \tilde H^*)(\tilde H^\dag\ {l}_{1L})}{M}
\left(- \frac{\phi_2^\dag\ \phi_1}{f^2} \right)^{4}\nn \\
&+&  \lambda_{2\nu} \left[
\frac{(\overline{{l}^c_{1L} }\ \tilde H^*)(\tilde H^\dag\ {L}_{L})}{M}
\frac{ i \sigma_2\
(\lambda_{1\nu}' \ \phi_1 - \lambda_{1\nu}'' \tilde\phi_2 )}{f}
\right.
 \nn \\
&& \quad \quad \quad
+   \left.
\frac{
(\lambda_{1\nu}' \ \phi_1 - \lambda_{1\nu}'' \tilde\phi_2)^T\ i \sigma_2^T }{f}
\frac{(\overline{{L}^c_{L} }\ \tilde H^*)(\tilde H^\dag\ {l}_{1L})}{M}
\right]
\left(- \frac{\phi_2^\dag\ \phi_1}{f^2}\right)
 \nn \\
&+&
\frac{
(\overline{{L}^c_{L} }\ \tilde H^*)^T
(i\sigma_2 \sigma_{\tau})
(\tilde H^\dag\ {L}_{L})^T
}{2 M_3 f^2}
\left[ - (\lambda_{3\nu}')^2 \ \phi_2^T i\sigma_2\sigma_{\tau} \phi_2
+ \lambda_{3\nu}' \lambda_{3\nu}''  \ \left({\tilde \phi_1}^T i\sigma_2\sigma_{\tau} \phi_2
+ { \phi_2}^T i\sigma_2\sigma_{\tau} \tilde\phi_1 \right) \right.
\nn \\
&& \quad
- \left. (\lambda_{3\nu}'')^2 \ \tilde\phi_1^T\ i\sigma_2\sigma_{\tau}\ \tilde\phi_1 \right] + H.c. \label{lnu}
\eea
where we have used a compact notation for the flavor doublets (transposition acts on flavor indices).

Analogously, for the charged leptons from \eq{yuklep} one obtains
\bea
- \mathcal{L}_{e} &=&
\overline{e_{R}}\ \Bigg[
- \lambda_{1e}\ (H^\dag\ {l}_{1L}) \left(-\frac{\phi_2^\dag \phi_1}{f^2} \right)^{3}\
+ (H^\dag\ {L}_{L})\frac{i\sigma_2 (\lambda_{2e}\ \phi_1  
- \lambda_{3e}\  \tilde\phi_2 )}{f}
 \Bigg]  \nn \\
&+&  \overline{E_{R}}^T \Bigg[
\frac{i\sigma_2 (\lambda_{1E}\ \phi_1  - \lambda_{1E}'\  \tilde\phi_2 )}{f} \left(-\frac{\phi_2^\dag \phi_1}{f^2} \right)^{2} \Bigg]
(H^\dag\ {l}_{1L})  \nn \\
&+&  \overline{E_{R}}^T \Bigg[
\lambda_{2E}  +  \lambda_{2E}'  \left(\frac{\phi_1^\dag \phi_1}{f^2} \right)
- \lambda_{2E}''  \left(\frac{\phi_2^\dag \phi_2}{f^2} \right)
%
+ \frac{\left(\lambda_{13E}\ \phi_1 \phi_1^\dag 
     - \lambda_{23E}'\ \tilde \phi_2 \tilde \phi_2^\dag + 
(1,+)\leftrightarrow (2,-)\right) }{f^2}  
\nn \\
&& \quad\quad\
- \frac{\left( \lambda_{14E}\ \phi_1 \tilde\phi_2^\dag 
- \lambda_{24E}'\ \tilde \phi_2 \phi_1^\dag + (1\leftrightarrow 2) \right)}{f^2}  
\Bigg] (H^\dag\ {L}_{L})^T
\  + H.c. \, , \label{le}
\eea
where transposition acts on flavor indices.

\subsection{Mass matrices}

When the  $SU(2)_F\times U(1)_F$ is broken, the little flavons assume their expectation values $v_1 \simeq v_2 \simeq \varepsilon f$ and we are left with the left-handed neutrino and charged-lepton mass matrices. At the lowest order in $\varepsilon$, factoring out a common coefficient $\lambda_{3\nu}^\prime$ in the leading terms and assuming all the other Yukawa couplings to be equal we obtain from \eq{lnu} the mass  matrix for the neutrinos
\be
{\cal M_\nu} = \lambda_{3\nu}^{\prime 2} \frac{\langle h_0 \rangle^2}{M_3} \varepsilon^2
\left( \begin{array}{c c c}   0 & 0& 0\\
 0 & 1 & 1\\
0 & 1  & 1  \\
\end{array} \right) \, . \label{nu-mass0}
\ee
In a similar manner, from the lagrangian (\ref{le}) after factoring out the coefficient $\lambda_{2E}$ of the leading terms in the charged-lepton matrix yields
\be
{\cal M}_l = \lambda_{2E} \langle h_0 \rangle
\left( \begin{array}{c c c}   0 & 0& 0\\0 & 1& 0\\0 & 0& 1 \\
\end{array} \right) \label{e-mass0} \, .
\ee

The coefficient
\be
 2 \, \lambda_{3\nu}^{\prime 2} \frac{\langle h_0 \rangle^2}{M_3} \varepsilon^2 = m_{\nu_\tau} \label{cond1}
  \ee
normalizes the absolute value of the matrix entries and fixes the value of the Yukawa coupling $\lambda_{3\nu}^\prime$.
The scale $M$  is just below or around $f$ and therefore we are not implementing the
usual see-saw mechanism that requires scales  as large as $10^{13}$ TeV. Therefore, neutrino masses are  in this model small because of the smallness of the corresponding effective Yukawa coupling $\lambda_{3\nu}^\prime$ that we take---for $M \simeq M_3 \simeq 10$ TeV---of the order of  $10^{-4}$.
This  value is necessary to give the correct absolute values for the square mass differences and it cannot be explained by the model; what the model explains is the hierarchy among the masses of different families.

We estimate the coefficient $\lambda_{2E}$ by means of value of the mass of the $\tau$ lepton in the relation
\be
\lambda_{2E}  \langle h_0  \rangle \simeq m_\tau \, , \label{cond2}
\ee
which yields a value for $\lambda_{2E}$ of about $10^{-2}$.

The form of the mass matrices thus obtained is  encouraging  because it shows a texture leading to normal hierarchy and maximal mixing in the neutrino sector as well as a first approximation to the charged-lepton masses. However, this result is not sufficient as it stands because the mixing angles and masses depends in a critical manner  on the exact values of all the entries. For instance, if we were to take the matrices (\ref{nu-mass0}) and (\ref{e-mass0})  as they stand, they would lead to a diagonal mixing matrix. While what we have found is a potentially correct texture, in order to reproduce the experimental data, it must have entries that are not equal even though they must, at the same time, only differ by $O(1)$ coefficients. This result is achieved by taking a closer look at the model. Before that, we pause to briefly comment on the problems of gauge anomalies and of the experimental bounds for the flavon masses.

 \subsection{Anomaly cancellation}

 Gauge anomalies are potentially present in the theory, as it can be easily seen by inspection considering the charges of the matter fields. They arise from  five kinds of triangle diagrams, that in the presence of leptons alone yields:
\bea
SU(2)_{F,EW}\times SU(2)_{F,EW}\times U(1)_F& \rightarrow &\Tr Y_F  = -4 \nn \\
U(1)_F\times U(1)_F\times U(1)_F & \rightarrow &\Tr Y_F^3 = -\frac{67}{4}\nn \\
U(1)_F\times U(1)_{EW}\times U(1)_{EW} & \rightarrow &\Tr Y_F\, Y_{EW}^2 =-\frac{5}{2} \nn \\
U(1)_F\times U(1)_F\times U(1)_{EW} & \rightarrow &\Tr Y_F^2\,Y_{EW} = -3 \nn
\eea
While a discussion of the anomalies cannot be done until also the quark fields are included and their charges with respect to the flavor gauge symmetries known, we notice that whatever anomaly is eventually found, it can be canceled  by a Wess-Zumino term. Since all the anomalous gauge symmetries are spontaneously broken, it is possible to build the required Wess-Zumino term by means of only the would-be-Goldstone bosons that are eventually eaten by the gauge fields. As shown in \cite{FPPS} this construction is sufficient to cancel all gauge anomalies (as well as those from gravity) without having to modify the matter content of the theory or strongly affecting the low-energy phenomenology.

\subsection{Bounds on the flavon masses}
 The most severe bounds on the little flavon interactions and masses,
in the context of the present discussion,
 come from  tightly constrained lepton-number violating
 processes like $\mu \rightarrow e \gamma$ and $\mu \rightarrow 3\,e$.
On the other hand, the inspection of the vertices involved in
diagrams with flavon exchange (lepton exchange can be safely
neglected) shows that a combination of loops, couplings and flavon
multiplicity severely suppresses the decay rates.

\begin{figure}
\begin{center}
\begin{picture}(400,70)(0,0)
\Vertex(250,35){1.5}
\Vertex(150,35){1.5}
\Text(99,15)[r b]{$e_R$}
\ArrowLine(150,35)(100,15)
\Text(99,55)[r t]{$\mu_L$}
\ArrowLine(100,55)(150,35)
\Text(200,23)[ b]{$\phi_{1,2}$}
\DashArrowLine(150,35)(250,35){2}
\Text(301,55)[l t]{$e_L$}
\ArrowLine(250,35)(300,55)
\Text(301,15)[l b]{$e_R$}
\ArrowLine(300,15)(250,35)
\Text(151,60)[]{$\times$}
\Text(153,50)[l]{$H$}
\DashLine(150,35)(150,60){1.5}
\Text(251,60)[]{$\times$}
\Text(253,50)[l]{$H$}
\DashLine(250,35)(250,60){1.5}
\Text(250,8)[]{$\times$}
\Text(250,0)[]{$\phi$}
\DashLine(250,35)(250,8){2}
\Text(270,15)[]{$\times$}
\DashLine(270,15)(250,35){2}
\Text(230,15)[]{$\times$}
\DashLine(230,15)(250,35){2}
\Text(260,8)[]{$\times$}
\DashLine(260,8)(250,35){2}
\Text(240,8)[]{$\times$}
\DashLine(240,8)(250,35){2}
\end{picture}
\end{center}
\caption{Flavon mediated contribution to the decay
$\mu\,\rightarrow\,e\,e\,e$}
\label{bounds}
\end{figure}
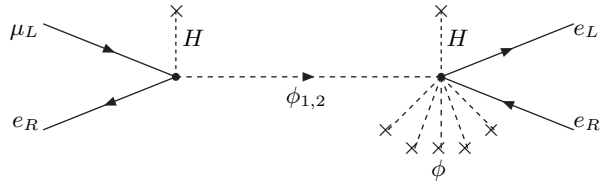

For instance, at
the tree level, we estimate from \eq{le} that the branching ratio for $\mu \to 3 e$ is (see \fig{bounds})
\be
BR(\mu \to 3 e) <
N^2 \left(\frac{\lambda_{1e}\lambda_{2e,3e}}{g_W^2}\right)^2\
\left(\frac{\vev{h_0}}{f}\right)^{4}\
\left(\frac{v_1 v_2}{f^2}\right)^{5}\
\left(\frac{m_W}{m_\phi}\right)^4 \simeq 10^{-10}\
\left(\frac{\vev{h_0}}{f}\right)^{4}\
\left(\frac{m_W}{m_\phi}\right)^4 \, ,
\label{muto3e}
\ee
where
$N$ is a combinatorial factor of order 10, the Yukawa couplings
$\lambda_{ie}$ are all of order $10^{-2}$ or less, and, as we shall
see in the numerical analysis, $v_1 v_2/f^2$ is of order
$10^{-1}$. The present experimental bound of $10^{-12}$ thus
allows for the existence of light flavons. In the range of scales
under consideration, inspection
of flavor gauge mediated contributions does not show the presence
of large effects either.

We
postpone a more detailed discussion of little flavon phenomenology to a
forthcoming paper.

\section{Lepton masses and mixing}

The more complete analysis requires that  the first non-vanishing entries in all  matrix elements be kept. In addition, we also retain $O(\varepsilon^2)$ corrections to all leading  entries---that means  $O(\varepsilon^4)$ terms in the neutrino mass matrix and $O(\varepsilon^2)$ terms in the charged lepton case.  The two vacua  are distinguished as $v_1=f \varepsilon_1$ and $v_2 = f \varepsilon_2$.  This analysis gives us the full textures.

Accordingly, the neutrino  Majorana mass matrix can now be written as
\begin{widetext}
\be
{\cal M_\nu} =  \frac{ \langle h_0 \rangle^2}{M} 
{\tiny
\left( \begin{array}{c c c}    \left[ r\ \lambda_{3\nu}^2 + 2 \lambda_{1\nu} \lambda_{2\nu} \right]   \varepsilon_{1}^4\varepsilon_{2}^4
& -\lambda_{2\nu} \lambda_{1\nu}' \varepsilon_{1}^{2} \varepsilon_{2}  & -\lambda_{2\nu}\lambda_{1\nu}'' \varepsilon_{1} \varepsilon_{2}^{2} \\
-\lambda_{2\nu} \lambda_{1\nu}'   \varepsilon_{1}^{2} \varepsilon_{2}
& r\ \lambda_{3\nu}'^2 \varepsilon_{2}^2 \rho + 2 \lambda_{1\nu}' \lambda_{2\nu}' \varepsilon_{1}^2 \varepsilon_{2}^2
& r\ \lambda_{3\nu}' \lambda_{3\nu}'' \varepsilon_{1} \varepsilon_{2} \rho
  +  \lambda_{1\nu}' \lambda_{2\nu}'' \varepsilon_{1}^3 \varepsilon_{2}
  +  \lambda_{2\nu}' \lambda_{1\nu}'' \varepsilon_{1}   \varepsilon_{2}^3  \\
-\lambda_{2\nu} \lambda_{1\nu}'' \varepsilon_{1} \varepsilon_{2}^{2}
&  r\ \lambda_{3\nu}' \lambda_{3\nu}''  \varepsilon_{1} \varepsilon_{2} \rho
  +   \lambda_{1\nu}' \lambda_{2\nu}''  \varepsilon_{1}^3 \varepsilon_{2}
  +  \lambda_{2\nu}' \lambda_{1\nu}'' \varepsilon_{1} \varepsilon_{2}^3
&r\ \lambda_{3\nu}''^2 \varepsilon_{1}^{2} \rho + 2 \lambda_{1\nu}'' \lambda_{2\nu}'' \varepsilon_{1}^2 \varepsilon_{2}^2 \\
\end{array} \right) } 
\label{neutral-mass} \, ,
\ee
\end{widetext}
where $\rho \equiv  1 - \varepsilon_1^2/3  - \varepsilon_2^2/3 $ comes from the expansion of the $\Sigma$ field.
The eigenvalues of this matrix are the masses of the three neutrinos.

In the same approximation, the Dirac mass matrix for the charged leptons is given by
\begin{widetext}
 \be
{\cal M}_l =   \langle h_0 \rangle
{\tiny
\left( \begin{array}{c c c}  \lambda_{1e}\, \varepsilon_{1}^3\varepsilon_{2}^3
& \lambda_{2e}\, \varepsilon_{1} & \lambda_{3e} \, \varepsilon_{2} \\
 \lambda_{1E} \,\varepsilon_{1}^{2} \varepsilon_{2}^3 
& \lambda_{2E} + (\lambda_{2E}'+ \lambda_{13E}')\  \varepsilon_1^2  
-(\lambda_{2E}'' + \lambda_{23E}')\ \varepsilon_2^2 & 
(\lambda_{14E}' + \lambda_{24E}')\ \varepsilon_1\varepsilon_2\  \\
 \lambda_{1E}' \,\varepsilon_{1}^{3}\ \varepsilon_{2}^{2} 
& - (\lambda_{14E} + \lambda_{24E})\ \varepsilon_1\varepsilon_2\ 
& \lambda_{2E} 
+(\lambda_{2E}'+ \lambda_{13E})\ \varepsilon_1^2
-(\lambda_{2E}'' + \lambda_{23E})\ \varepsilon_2^2  \\
\end{array} \right) }
\label{charged-mass} \, .
\ee
\end{widetext}

The $3\times3$ unitary PMNS mixing matrix~\cite{PMNS} for the leptons is defined as
\be
U = U_{l}^\dag U_{\nu}
\ee
where $U_{\nu},\,U_{l}$ are the neutrino and charged lepton mixing matrices defined, respectively, by
\be
U_l^\dag {\cal M}_l^\dag {\cal M}_l U_l = \left( {\cal M}_l^D \right)^2
\quad \mbox{and} \quad U_\nu^\dag {\cal M}_\nu U_\nu = {\cal M}_\nu^D \, ,
\ee
where $ {\cal M}_l^D$ and $ {\cal M}_\nu^D$ are the diagonal mass matrices for, respectively, charged and neutral leptons.

We use for $U$ the standard parametrization (in the case of a real matrix)
\begin{widetext}
\be
U = \left( \begin{array}{c c c}   c_{12} c_{13} & c_{13} s_{12}  & s_{13}  \\
 -c_{23} s_{12} - c_{12} s_{12} s_{23} & c_{12} c_{23} - s_{12} s_{13} s_{23} & c_{13} s_{23}  \\
 -c_{12} c_{23} s_{13} + s_{12} s_{23}& -c_{23} s_{12} s_{13} - c_{12} s_{23}  &   c_{13} c_{23}\\
\end{array} \right) \, ,\label{pmns}
\ee
\end{widetext}
where $s_{ij} = \sin \theta_{ij}$ and $c_{ij} = \cos \theta_{ij}$ and thus obtain
 a $U$ parametrized by  three angles.

The mass matrices (\ref{neutral-mass}) and (\ref{charged-mass}) necessarily contain many parameters. The  coupling strengths in front of the various Yukawa terms of the lagrangian can be different  for different flavors and different interactions and we have kept them distinguished so far.  For the model to be natural,  these parameters must  be roughly of the same order once their overall values have been fixed by  \eq{cond1} and (\ref{cond2}) respectively.

\subsection{Experimental data}

Let us briefly review the experimental results. Compelling evidences in favor of neutrino oscillations and, accordingly of non-vanishing neutrino masses has been collected in recent years from neutrino experiments~\cite{experiments}. Combined analysis of the experimental data show that the neutrino mass matrix is characterized by a hierarchy with two square mass differences (at $99.73\%$ c.l.):
\bea
\Delta m^2_\odot & = & (3 - 35) \times 10^{-5} \mbox{eV}^2 \nn\\
|\Delta m^2_\oplus | &= & (1.4 - 3.7) \times 10^{-3} \mbox{eV}^2 \, , \label{delta-masses}
\eea
the former controlling solar neutrino oscillations~\cite{Sandhya} and the latter the atmospheric neutrino experiments~\cite{Fogli}.
In the context of three active neutrino oscillations, the mixing
is described by the
PMNS mixing matrix $U$ in \eq{pmns}.
Such a matrix is parameterized by three mixing angles, two of which can be identified with the mixing angles determining solar~\cite{Sandhya} and atmospheric~\cite{Fogli} oscillations, respectively (again, at $99.73\%$ c.l.):
 \bea
 \tan^2 \theta_\odot & = &  0.25 - 0.88 \, ,  \nn \\
 \sin^2 2\, \theta_\oplus  & = & 0.8 - 1.0 \, .  \label{mix-angles}
 \eea

For the third angle, controlling the mixing $\nu_\tau$-$\nu_e$, there are at present only upper limits, deduced by reactor neutrino experiments~\cite{reactors} (at 95\% c.l.):
\be
\sin^2  \theta < 0.09 \, .
\ee

Finally, the charged-lepton masses are well known and given by
$m_\tau\simeq 1770$ MeV, $m_\mu\simeq 106$ MeV and  $m_e
 \simeq 0.5$ MeV, respectively---so that, $m_\tau/m_\mu \simeq 17$ and $m_\mu/m_e \simeq 207$.

\subsection{Numerical results}

The goal of our numerical analysis is to show that, in spite of the
many undetermined couplings in \eq{neutral-mass} and
\eq{charged-mass}, the known pattern of mixing angles and
neutrino mass differences is rather well reproduced by setting
all the  ratios of Yukawa couplings---remaining after extracting the overall factors in \eqs{nu-mass0}{e-mass0}---equal to 1. The textures induced by
the structure of the flavon VEV's already determine the desired
result. Just considering variation of $O(1)$ of a few couplings
allows for the complete fit of all lepton masses. We argue
therefore that the model reproduces quite naturally the known
structure of the leptonic spectrum.


Let us then---after having fixed the overall factors by \eqs{cond1}{cond2}---keep as input parameters of the theory the vacuum
values $v_1 = \varepsilon_1 f$ and $v_2 = \varepsilon_2 f$ while
taking all ratios of Yukawa couplings and $r$ equal in modulus to 1. This
procedure leaves us with only two parameters in terms of which we
may write the following toy mass matrices:
 \begin{widetext}
\be
{\cal M_\nu} = \lambda_{3\nu}^{\prime 2} \frac{\langle h_0 \rangle^2}{M_3}
\left( \begin{array}{c c c}   3\, \varepsilon_{1}^4\varepsilon_{2}^4
& - \varepsilon_{1}^{2} \varepsilon_{2}  & -  \varepsilon_{1} \varepsilon_{2}^{2} \\
- \varepsilon_{1}^{2} \varepsilon_{2}
& \varepsilon_{2}^2 \rho + 2 \, \varepsilon_{1}^2 \varepsilon_{2}^2
&  \varepsilon_{1} \varepsilon_{2} \rho
  +   \varepsilon_{1}^3 \varepsilon_{2}
  +  \varepsilon_{1}   \varepsilon_{2}^3   \\
- \varepsilon_{1} \varepsilon_{2}^{2}
&   \varepsilon_{1} \varepsilon_{2} \rho
  +       \varepsilon_{1}^3 \varepsilon_{2}
  +  \varepsilon_{1} \varepsilon_{2}^3
& \varepsilon_{1}^{2} \rho + 2   \varepsilon_{1}^2 \varepsilon_{2}^2 \\
\end{array} \right) \label{nu-mass-maj}
\ee
\end{widetext}
and
\begin{widetext}
 \be
{\cal M}_l = \lambda_{2E}  \langle h_0 \rangle
\left( \begin{array}{c c c}   \varepsilon_{1}^3\varepsilon_{2}^3 &  \varepsilon_{1}  &  \varepsilon_{2}  \\
\varepsilon_{1}^{2} \varepsilon_{2}^3 & 
1  & 
2 \varepsilon_{1}\varepsilon_2  \\
\varepsilon_{1}^{3} \varepsilon_{2}^{2} & 
2 \varepsilon_{1}\varepsilon_2 & 
1  \\
\end{array} \right) \label{nu-mass-dirac} \, .
\ee
\end{widetext}

  At this point we can vary our two parameters to  find the best fit.
We take the range for $ \varepsilon_1$  and
  $ \varepsilon_2$ between 0.1 and 1 in such a way that we do not introduce
large VEV's hierarchies.
 As an example,   for  the representative values:
 $$
 \varepsilon_1 \simeq  0.1 \quad \mbox{and} \quad \varepsilon_2\simeq 0.8\, ,
 $$
 we obtain for the mixing angles
  \bea
 \tan^2  \theta_\odot \; & \simeq & 0.9\, , \nn  \\
   \sin ^2 2\, \theta_\oplus & \simeq & 1\, ,\\
  \sin^2 \theta \quad & \simeq &  0.006\, ,  \nn
 \eea
for the neutrino normal-hierarchy ratio
\be
\Delta m^2_{12} /  \Delta m^2_{23}  \simeq 0.005 \, ,
\ee
and for the  charged-lepton mass hierarchy
 \be
m_\tau / m_\mu  \simeq 1.6 \quad \mbox{and} \quad m_\tau / m_e \simeq 3470  \, .
\ee

Albeit using a very rough approximation, we find
values for the mixing angles and neutrino masses in the ballpark of the experimental values, with the exception of the 
$\mu$ mass;
this is  to be expected since $\mu$ and $\tau$  start out as 
members of a $SU(2)_F$ doublet and their splitting must come from the detailed
values of the Yukawa couplings.
As a matter of fact, in order to fit correctly all data,
it is enough to
consider the case in which some of the Yukawa couplings  in the Dirac mass matrix (\ref{charged-mass}) differ by $O(1)$ coefficients.
We have checked that this is possible, as a matter of fact, with just two more parameters.
Therefore, without introducing any large ratio among Yukawa couplings
or VEV's,
all the known experimental data for the lepton masses and mixing are
reproduced.


 \acknowledgments

It is a pleasure to thank M.\ Frigerio, E.\ Pallante and S. Petcov
for discussions. One of us (MP) also thanks T. Appelquist and
W.\ Skiba for discussions and SISSA for the hospitality.
This work is
partially supported by  the European TMR Networks HPRN-CT-2000-00148
and HPRN-CT-2000-00152. The work of MP is supported in part by the
US Department of Energy under contract DE-FG02-92ER-40704.



\begin{thebibliography}{99}

\bibitem{littlehiggs}

N.~Arkani-Hamed, A.~G.~Cohen, E.~Katz and A.~E.~Nelson,
JHEP {\bf 0207}, 034 (2002)
[arXiv:hep-ph/0206021];

I.~Low, W.~Skiba and D.~Smith,
Phys.\ Rev.\ D {\bf 66}, 072001 (2002)
[arXiv:hep-ph/0207243];

D.~E.~Kaplan and M.~Schmaltz,
arXiv:hep-ph/0302049.

S.~Chang and J.~G.~Wacker,
arXiv:hep-ph/0303001.

W.~Skiba and J.~Terning,
arXiv:hep-ph/0305302.

S.~Chang,
arXiv:hep-ph/0306034.




\bibitem{flavormodels-old}

H.~Harari, H.~Haut and J.~Weyers,
Phys.\ Lett.\ B {\bf 78}, 459 (1978).

C.~D.~Froggatt and H.~B.~Nielsen,
Nucl.\ Phys.\ B {\bf 147}, 277 (1979).


T.~Maehara and T.~Yanagida,
Prog.\ Theor.\ Phys.\  {\bf 61}, 1434 (1979).

G.~B.~Gelmini, J.~M.~Gerard, T.~Yanagida and G.~Zoupanos,
Phys.\ Lett.\ B {\bf 135}, 103 (1984).



\bibitem{flavormodels-new} A partial, and by no means complete  list includes the following works:


M.~Dine, R.~G.~Leigh and A.~Kagan,
Phys.\ Rev.\ D {\bf 48}, 4269 (1993)
[arXiv:hep-ph/9304299].

M.~Leurer, Y.~Nir and N.~Seiberg,
Nucl.\ Phys.\ B {\bf 398}, 319 (1993)
[arXiv:hep-ph/9212278];
Nucl.\ Phys.\ B {\bf 420}, 468 (1994)
[arXiv:hep-ph/9310320].

P.~Pouliot and N.~Seiberg,
Phys.\ Lett.\ B {\bf 318}, 169 (1993)
[arXiv:hep-ph/9308363].

D.~B.~Kaplan and M.~Schmaltz,
Phys.\ Rev.\ D {\bf 49}, 3741 (1994)
[arXiv:hep-ph/9311281].

L.~J.~Hall and H.~Murayama,
Phys.\ Rev.\ Lett.\  {\bf 75}, 3985 (1995)
[arXiv:hep-ph/9508296].

A.~Pomarol and D.~Tommasini,
Nucl.\ Phys.\ B {\bf 466}, 3 (1996)
[arXiv:hep-ph/9507462];

R.~Barbieri, G.~R.~Dvali and L.~J.~Hall,
Phys.\ Lett.\ B {\bf 377}, 76 (1996)
[arXiv:hep-ph/9512388];

P.~H.~Frampton and O.~C.~Kong,
Phys.\ Rev.\ Lett.\  {\bf 77}, 1699 (1996)
[arXiv:hep-ph/9603372].

E.~Dudas, C.~Grojean, S.~Pokorski and C.~A.~Savoy,
Nucl.\ Phys.\ B {\bf 481}, 85 (1996)
[arXiv:hep-ph/9606383].


P.~Binetruy, S.~Lavignac and P.~Ramond,
Nucl.\ Phys.\ B {\bf 477}, 353 (1996)
[arXiv:hep-ph/9601243].



R.~Barbieri, L.~J.~Hall, S.~Raby and A.~Romanino,
Nucl.\ Phys.\ B {\bf 493}, 3 (1997)
[arXiv:hep-ph/9610449];

G.~Altarelli and F.~Feruglio,
Phys.\ Rept.\  {\bf 320}, 295 (1999).

H.~Fritzsch and Z.~z.~Xing,
Prog.\ Part.\ Nucl.\ Phys.\  {\bf 45}, 1 (2000)
[arXiv:hep-ph/9912358].


Z.~Berezhiani and A.~Rossi,
Nucl.\ Phys.\ B {\bf 594}, 113 (2001)
[arXiv:hep-ph/0003084];

A.~Masiero, M.~Piai, A.~Romanino and L.~Silvestrini,
Phys.\ Rev.\ D {\bf 64}, 075005 (2001)
[arXiv:hep-ph/0104101].


M.~Frigerio and A.~Y.~Smirnov,
Nucl.\ Phys.\ B {\bf 640}, 233 (2002)
[arXiv:hep-ph/0202247].


\bibitem{coleman-weinberg}
S.~R.~Coleman and E.~Weinberg,
Phys.\ Rev.\ D {\bf 7}, 1888 (1973).

 \bibitem{wip} F.~Bazzocchi \etal , to appear.


\bibitem{PMNS} B. Pontecorvo, Sov.\ Phys.\ JETP {\bf 6} (1958) 429;\\
Z. Maki, M. Nakagawa and S. Sakata, Prog.\ Theor.\ Phys.\ {\bf 28} (1962) 870.




\bibitem{LSS} I. Low, W. Skiba and D. Smith , in ~\cite{littlehiggs}

\bibitem{sher}
M.~Sher,
Phys.\ Rept.\  {\bf 179}, 273 (1989).

\bibitem{FPPS}
M.~Fabbrichesi, R.~Percacci, M.~Piai and M.~Serone,
Phys.\ Rev.\ D {\bf 66}, 105028 (2002)
[arXiv:hep-th/0207013].

\bibitem{seesaw} T. Yanagida, in \textit{Unified Theory and Baryon Number}, Tsukuba, 1979;

M. Gell-Mann, P. Ramond and R. Slansky, in \textit{Supergravity}, edited by P. van Nieuwehuizen and O. Freedman (North-Holland, Amsterdam 1979), p. 317;

R.~N.~Mohapatra and G.~Senjanovic,
Phys.\ Rev.\ Lett.\  {\bf 44}, 912 (1980).

\bibitem{tom}
T.~Appelquist and R.~Shrock,
Phys.\ Lett.\ B {\bf 548}, 204 (2002)
[arXiv:hep-ph/0204141].

\bibitem{lindner}
K.\ R.\ S.\ Balaji, M.\ Lindner and G.\ Seidl,  arXiv:hep-ph//0303245.

 \bibitem{experiments}
Y.~Fukuda {\it et al.}  [Super-Kamiokande Collaboration],
Phys.\ Rev.\ Lett.\  {\bf 81}, 1562 (1998) [arXiv:hep-ex/9807003];

S.~Fukuda {\it et al.}  [Super-Kamiokande Collaboration],
Phys.\ Rev.\ Lett.\  {\bf 86}, 5656 (2001) [arXiv:hep-ex/0103033];

S.~Fukuda {\it et al.}  [Super-Kamiokande Collaboration],
Phys.\ Rev.\ Lett.\  {\bf 86}, 5651 (2001) [arXiv:hep-ex/0103032];

Q.~R.~Ahmad {\it et al.}  [SNO Collaboration],
Phys.\ Rev.\ Lett.\  {\bf 87}, 071301 (2001)
[arXiv:nucl-ex/0106015];

Q.~R.~Ahmad {\it et al.}  [SNO Collaboration],
Phys.\ Rev.\ Lett.\  {\bf 89}, 011301 (2002)
[arXiv:nucl-ex/0204008];

M.~H.~Ahn {\it et al.}  [K2K Collaboration],
Phys.\ Rev.\ Lett.\  {\bf 90}, 041801 (2003)
[arXiv:hep-ex/0212007];

K.~Eguchi {\it et al.}  [KamLAND Collaboration],
Phys.\ Rev.\ Lett.\  {\bf 90}, 021802 (2003)
[arXiv:hep-ex/0212021].

\bibitem{Sandhya}
See, for instance: S.~Choubey, A.~Bandyopadhyay, S.~Goswami and D.~P.~Roy,
arXiv:hep-ph/0209222.

\bibitem{Fogli}
See, for instance: G.~L.~Fogli, E.~Lisi, A.~Marrone and D.~Montanino,
arXiv:hep-ph/0303064.

\bibitem{reactors}
M.~Apollonio {\it et al.}  [CHOOZ Collaboration],
Phys.\ Lett.\ B{\bf 466} (1999) 415 (hep-ex/9907037);

F. Boehm, J. Busenitz et al., Phys.\ Rev.\ Lett.\  {\bf 84}
(2000) 3764 and Phys. Rev. D{\bf 62} (2000) 072002.


 \end{thebibliography}
 \end{document}